\documentclass[aps,pra,reprint,longbibliography,superscriptaddress,nofootinbib]{revtex4-1}

\usepackage{amsmath,amssymb,MnSymbol,graphicx,amsfonts}

\begin{document}

\title{Ultrasensitive Inverse Weak-Value Tilt Meter}

\author{Juli\'{a}n Mart\'{i}nez-Rinc\'{o}n}
\email{jmarti41@ur.rochester.edu}
\affiliation{Department of Physics and Astronomy, University of Rochester, Rochester, New York 14627, USA}
\affiliation{Center for Coherence and Quantum Optics, University of Rochester, Rochester, New York 14627, USA}
\author{Christopher A. Mullarkey}
\affiliation{Department of Physics and Astronomy, University of Rochester, Rochester, New York 14627, USA}
\affiliation{Center for Coherence and Quantum Optics, University of Rochester, Rochester, New York 14627, USA}
\author{Gerardo I. Viza}
\email{Present Address: Intel Corporation, Hillsboro, Oregon 97124, USA}
\affiliation{Department of Physics and Astronomy, University of Rochester, Rochester, New York 14627, USA}
\affiliation{Center for Coherence and Quantum Optics, University of Rochester, Rochester, New York 14627, USA}
\author{Wei-Tao Liu}
\affiliation{Department of Physics and Astronomy, University of Rochester, Rochester, New York 14627, USA}
\affiliation{Center for Coherence and Quantum Optics, University of Rochester, Rochester, New York 14627, USA}
\affiliation{College of Science, National University of Defense Technology, Changsha, 410073, China}
\affiliation{Interdisciplinary Center of Quantum Information, National University of Defense Technology, Changsha, 410073, China}
\author{John C. Howell}
\affiliation{Department of Physics and Astronomy, University of Rochester, Rochester, New York 14627, USA}
\affiliation{Center for Coherence and Quantum Optics, University of Rochester, Rochester, New York 14627, USA}
\affiliation{Institute of Optics, University of Rochester, Rochester, New York 14627, USA}
\affiliation{Institute for Quantum Studies, Chapman University, Orange, California 92866, USA}




\begin{abstract}
We present an interferometric technique for measuring ultra-small tilts. The information of a tilt in one of the mirrors of a modified Sagnac interferometer is carried by the phase difference between the counter propagating laser beams. Using a small misalignment of the interferometer, orthogonal to the plane of the tilt, a bimodal (or two-fringe) pattern is induced in the beam's transverse power distribution. By tracking the mean of such a distribution, using a split detector, a sensitive measurement of the phase is performed. With 1.2 mW of continuous-wave laser power, the technique has a shot noise limited sensitivity of 56 frad/$\sqrt{\mbox{Hz}}$, and a measured noise floor of 200 frad/$\sqrt{\mbox{Hz}}$ for tilt frequencies above 2 Hz. A tilt of 200 frad corresponds to a differential displacement of 4.0 fm in our setup. The novelty of the protocol relies on signal amplification due to the misalignment, and on good performance at low frequencies. A noise floor of about 70 prad/$\sqrt{\mbox{Hz}}$ is observed between 2 and 100 mHz.
\end{abstract}



\maketitle
Precision measurements of an angular deflection or tilt are crucial in different areas of engineering and science. Ultra-small-tilt sensors are used, to name a few, in atomic-force microscopy~\cite{AFM}, in alignments of the LIGO configuration using optical levers~\cite{LIGO}, and in measurements of angular rotations in torsion pendula~\cite{ReviewofG} and torsion-bar antennas~\cite{TorsionAntenna}. 

Interferometric designs offer great performance for tilt measurements~\cite{LISA,Patent1,Patent2}. Using a Mach-Zehnder configuration, Park and Cho~\cite{ParkTilt} recently reported a noise-floor of 10 prad/$\sqrt{\mbox{Hz}}$ at a frequency slightly above 1 Hz and of 400 frad/$\sqrt{\mbox{Hz}}$ for frequencies above 30 Hz. Weak-value amplification techniques in Sagnac configurations have also been explored~\cite{Dixon,Starling,Turner,Kasevich,Concatenated,Weitao}. These protocols offer technical-noise mitigation advantages~\cite{PRX,NoiseExp} and rely only on one split detector instead of elaborated detection designs (see for example Ref.~\cite{ParkTilt}). These weak-values based techniques transfer the tilt information into a phase ramp in the interferometer, and make use of a tunable constant phase to amplify the signal via post-selection. On the other hand, inverse-weak-value amplification~\cite{inverse} offers the possibility to encode the desired information in a constant phase instead of a phase ramp in the interferometer. Such a protocol has been unexplored for precision measurements by the scientific community, and should not be confused with the better known technique of weak-value amplification~\cite{Dixon,Starling,Turner,Kasevich,Concatenated,Hosten,David,Velocimetry,LuisJoseDelays,LuisJose,ModifiedHallEffect,HallEffectNanometal,Jayaswal,Pfeifer,LuisJoseTemperature,Thermostat,AngularMomentum} and the recently developed technique of almost-balanced weak values~\cite{BWV,Weitao,Zekai}.  

We  present in this letter the use of the technique of inverse-weak-value amplification in a modified Sagnac configuration~\cite{SagnacDesign}. We improve the sensitivity for high-frequency tilt measurements on table-top interferometric setups, and expand the field to the mHz regime. The used interferometric Sagnac configuration, with the two paths spatially separated, allows for a lower shot-noise bound than the case of a collinear configuration~\cite{Weitao,Starling}. A similar advantage was also used in Ref~\cite{ParkTilt}. 

\begin{figure}
\centering
\includegraphics[width=\linewidth]{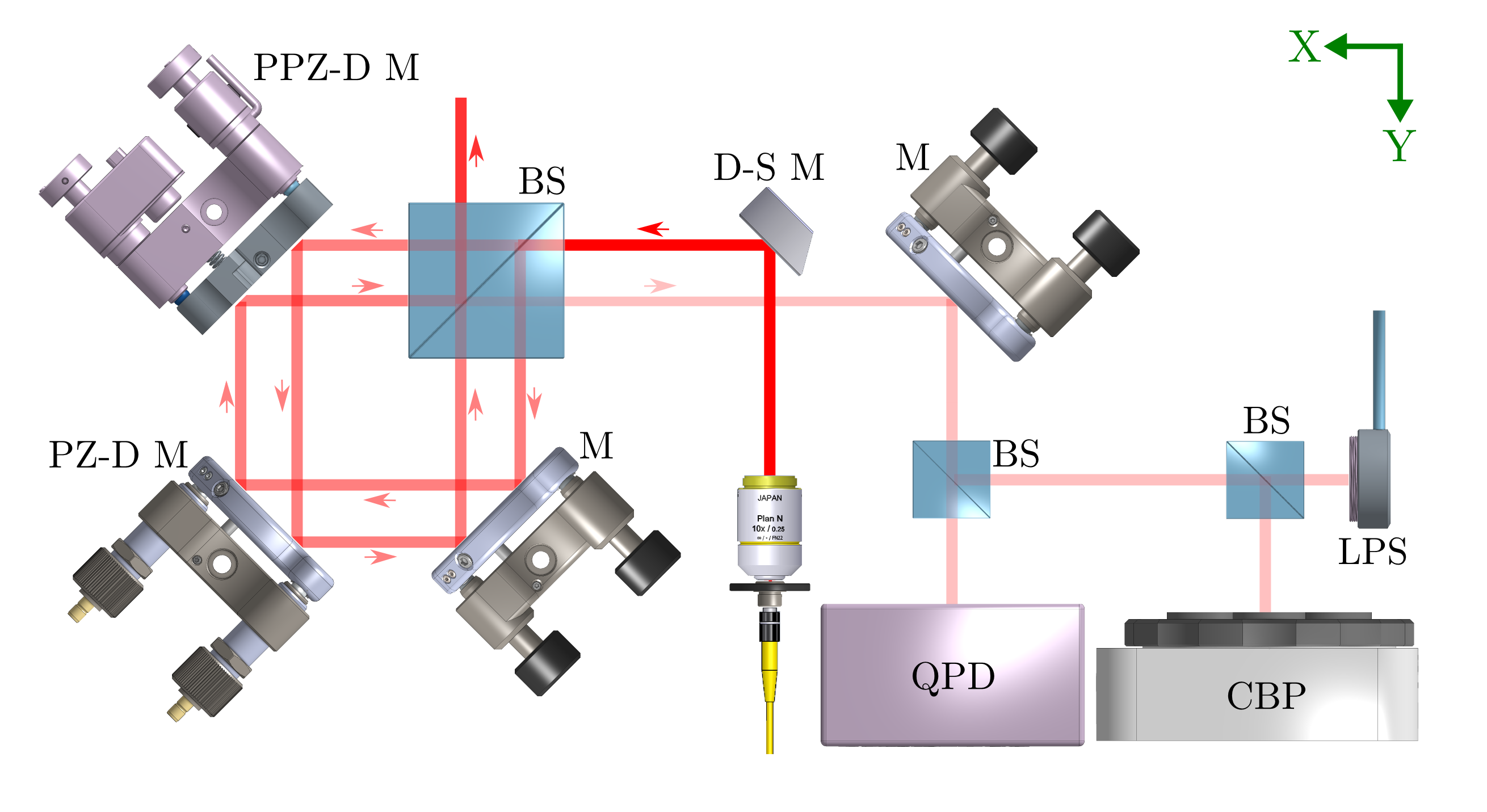}
\caption{Experimental setup to measure a mirror's tilt. A laser beam enters a modified Sagnac interferometer through a beam splitter (BS) such that the two counter propagating paths are spatially separated. A sinusoidal small tilt on the x-y plane is induced using a piezo-driven mirror (PZ-D M), and a larger and constant misalignment tilt in the perpendicular direction is applied using a picomotor piezo-driven mirror (PPZ-D M). The mean of the beam's transverse power distribution at the dark port is measured using a quadrant position detector (QPD). The bi-modal transverse distribution of the beam in the z-direction is tracked using a camera beam profiler (CBP) and a laser power sensor (LPS). M: Mirror, D-S M: D-Shaped mirror.}
\label{fig:setup}
\end{figure}

A TEM00 laser beam is used to measure a ultra small tilt induced by a piezo-driven mirror (PZ-D M in Fig.~\ref{fig:setup}, Thorlabs Polaris-K1PZ) in a modified Sagnac interferometer. The tilt information is mapped onto the relative phase between the paths of the interferometer by making the two counter-propagating beams spatially separated. The induced relative phase is given by $\phi=\sqrt{2}k_0L\theta$, where $\theta$ is the tilt of the mirror in the x-y plane, $L$ is the separation of the beams at the surface of the tilted mirror, and $k_0$ is the wave number. A larger relative phase ramp in the z-direction is induced by slightly misaligning the interferometer using a picomotor piezo-driven mirror (PPZ-D M in Fig.~\ref{fig:setup}, Newport 8821). After hitting the misalignment mirror both beams propagate different distances in the interferometer before recombining again at the output. The z-component of the transverse field intensity at the dark port takes the form $I_{out}(z)\propto |1-e^{i(\phi+kz)}|^2e^{-z^2/2\sigma^2}$,
where $k$ is the effective transverse momentum kick inducing the misalignment and $2\sigma$ is the $1/e^{2}$ radius of the input Gaussian beam. 

\begin{figure}
\centering
\includegraphics[width=0.8\linewidth]{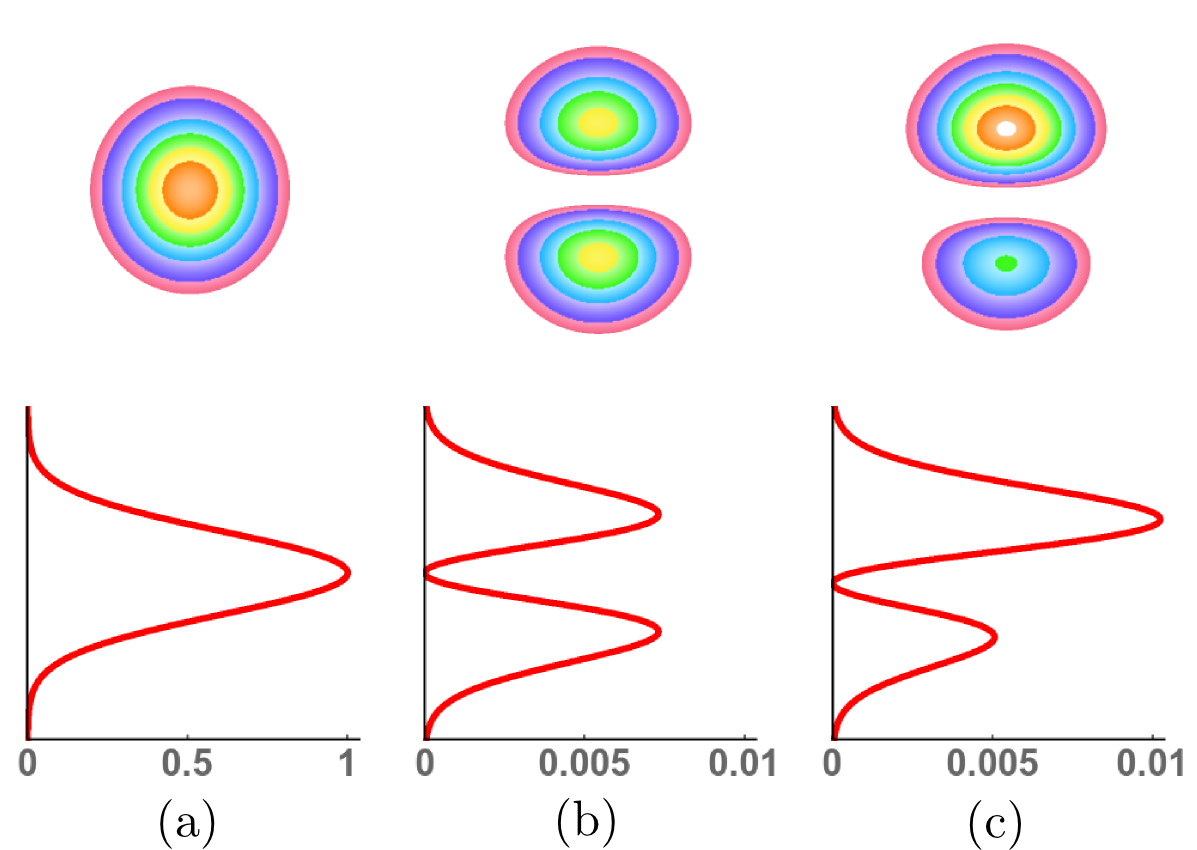}
\caption{Simulations of beam profiles (upper) and power distributions in the z-direction (bottom). Figure (a) shows the input TEM00 mode, (b) shows the balanced bimodal distribution at the output when $\phi=0$, and (c) shows the nonsymmetric distribution for $\phi=0.05$. $k\sigma=0.2$ in (b) and (c), so the profiles (upper) are increased 100 times in intensity with respect to (a) for visualization purposes.}
\label{fig:density}
\end{figure}

The system is set satisfying the condition $\phi\ll k\sigma\ll1$, such that a transverse bimodal pattern is seen at the output. If $\phi=0$ a balanced distribution is obtained such that $\langle z\rangle$=0, and a small phase $\phi$ breaks the symmetry. See Fig.~\ref{fig:density} for a simulation of such output distribution. The novelty of the technique relies on tracking the mean of such a distribution using a split detector. Note that conventional weak-value techniques consist on measuring a shift of shape-preserved distributions~\cite{Dixon,Turner,Kasevich,PRX,NoiseExp,Concatenated}, and standard interferometric techniques consist on measuring power changes at the optical outputs (measurements of the number of fringes in a dark port, Homodyne detection, etc.).   

The mean of the distribution takes the form
\begin{equation}\label{shift}
\langle z\rangle=\frac{\int_{-\infty}^{\infty}z\cdot I_{out}\,dz}{\int_{-\infty}^{\infty}I_{out}\,dz}=\frac{k\sigma^2\sin\phi}{e^{k^2\sigma^2/2}-\cos\phi}\approx\frac{2\phi}{k},
\end{equation}
meaning that the smaller the misalignment $k$ the larger the shift of the mean, facilitating the estimate of $\phi$. This result is a common amplification response of weak-values techniques. Also, the larger the amplification of the shift $\langle z\rangle$ the lesser the detected laser power, which is approximately $(k\sigma/2)^2$ smaller than the total input power to the interferometer. 

If $N$ photons are sent to the interferometer, the shot noise, using a split detector~\cite{splitdetector}, when estimating the tilt is given by 

\begin{equation}\label{shot-noise}
\Delta\theta_{SN}=\frac{\Delta\phi_{SN}}{\sqrt{2}k_0L}
=\frac{k}{2\sqrt{2}k_0L}\times\frac{\sqrt{\pi/2}\,\sigma}{\sqrt{N(k\sigma/2)^2}}
=\frac{1}{4\sqrt{\pi}}\left(\frac{\lambda}{L\sqrt{N}}\right).
\end{equation}

This lower bound for sensitivity in estimates of the tilt $\theta$ is smaller by a factor of $\sqrt{2}\sigma/L$ with respect to the equivalent weak-value amplification technique in a collinear Sagnac configuration~\cite{Weitao,Starling}. The fact that the separation of the beams, $L/\sqrt{2}$, can be much larger than the beam radius, 2$\sigma$, is the principal advantage of using the modified Sagnac configuration.

\begin{figure}
\centering
\includegraphics[width=0.9\linewidth]{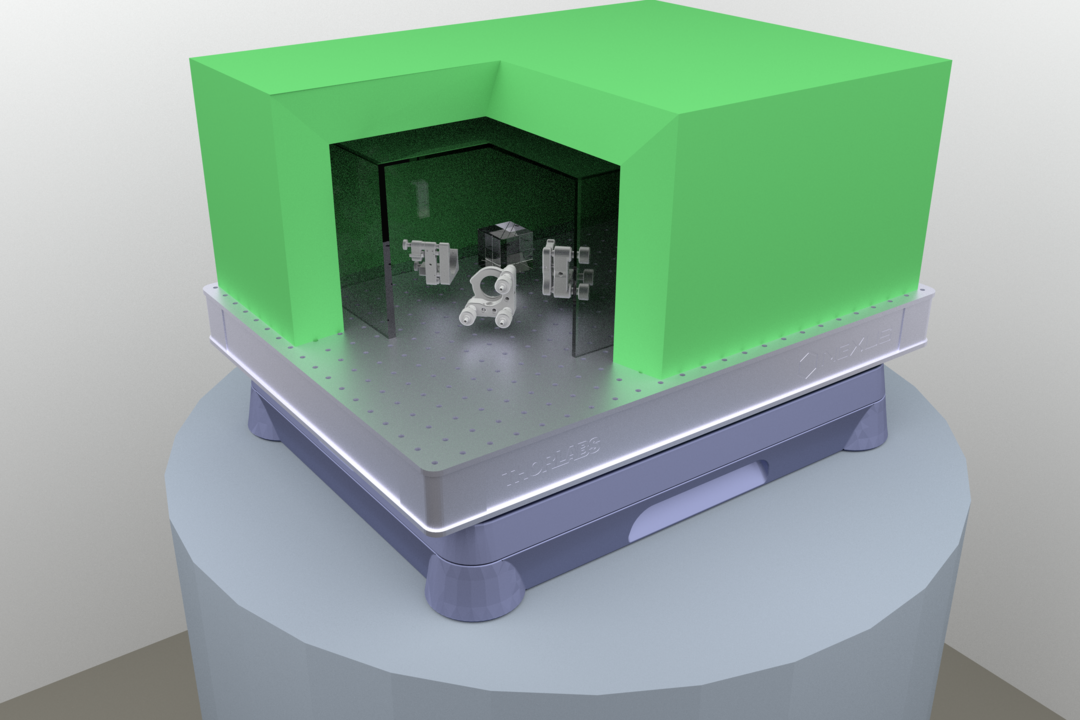}\caption{Three-dimensional schematic of the experimental design. The optical configuration is enclosed using a plastic cover and insulation foam, which are cut away in the drawing to show the interior of the design. No all optical elements are shown.}
\label{fig:3D}
\end{figure}

We built the optical setup, as shown in Fig.~\ref{fig:setup}, in an area of approximate 8"x17" on top of a 18"x24" optical breadboard (Thorlabs B1824F), as shown in Fig.~\ref{fig:3D}. A plastic cover and 2"-thick insulation foam were used to avoid air currents and drastic thermal fluctuations. We note that no active control (cooling or heating) of the temperature was present in the room where the experiment was performed. A 780 nm laser (New Focus Tunable Diode Laser 7000) was fiber-coupled far outside from the setup to avoid constant heating. 
1.2 mW of continuous-wave power was fed to the interferometer through a single-mode optical fiber. 
The system was put on top of an active vibration cancellation platform (TMC Everstill K-400) to improve stability in the 1-10 Hz range. In addition, the whole system was on top of a large concrete pillar resting on sand and isolated from the building. The output differential electrical signal of the split detector (upper minus lower quadrants of the QPD in Fig.~\ref{fig:setup}, Newport 2901) was sent to two, connected-in-series, low-noise voltage preamplifiers (Standard Research Systems SR560) for low-pass frequency filtering and amplification. The output signal was recorded using an oscilloscope connected to a computer. The beam's diameter ($4\sigma$) was about 1 mm. The total detected power was about 8.8 times smaller than the power detected when the output was set at the bright port, i.e. $(k\sigma/2)^2\sim0.1$. This means that the misalignment angle was about $k/k_0\sim0.3$ mrad. 

\begin{figure*}
\centering
\includegraphics[width=0.99\linewidth]{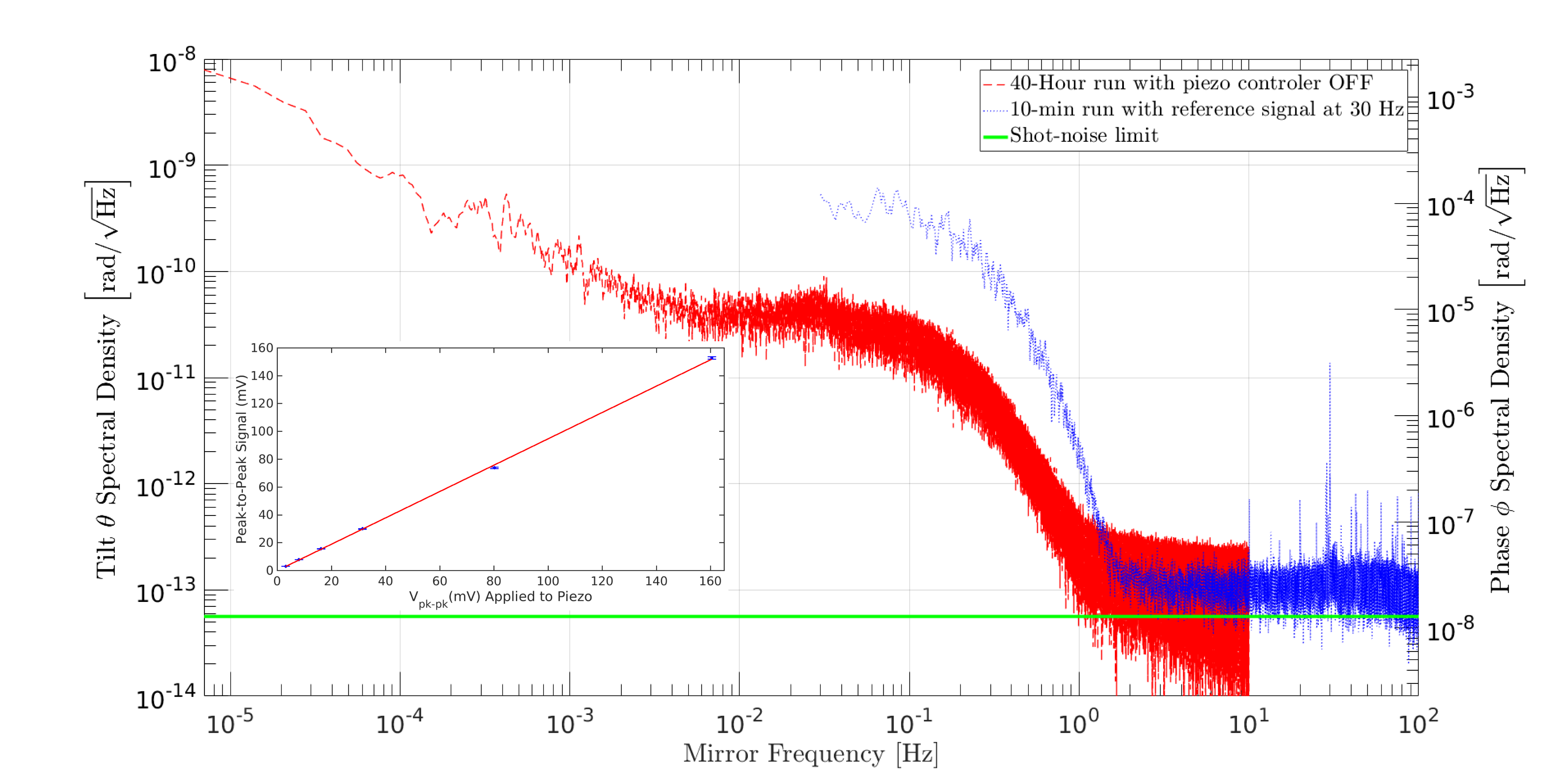}
\caption{Mirror's tilt (left scale) and relative phase in the interferometer (right scale) spectral densities. The dotted blue curve corresponds to the FFT of a 10-minute run with a reference tilt signal of 1.6 nrad at 30 Hz. The dashed red curve corresponds to the FFT of a 40-hour run without any reference signal to avoid electronic noise caused by the function generator and the piezo controller. Inset: Linear response of the tilt sensor for different applied voltages to the piezo actuator.}
\label{fig:spectrum}
\end{figure*}

Figure~\ref{fig:spectrum} shows the result of the measurements. First, a sinusoidal signal of 321 $\mu$V peak-to-peak at 30 Hz was applied to the piezo actuator, inducing a 1.6 nrad peak-to-peak tilt in the mirror. The manual-given value of 5.0 $\mu$rad/V was used for the piezo response. The output differencing signal was filtered using low-noise preamplifiers as 6 dB/oct rolloff band-pass filters between 30 mHz and 300 Hz, and with 100-folded amplification. A 10-minute raw signal was recorded on the computer using a 1 kHz acquisition rate on the oscilloscope. The Fast Fourier Transform (FFT) of the time-signal is shown in dotted blue in Fig.~\ref{fig:spectrum}. The technique is nearly shot-noise limited for frequencies larger than 2 Hz, having a noise floor of about 200 frad/$\sqrt{\mbox{Hz}}$. The shot-noise ($\sim$56 frad/$\sqrt{\mbox{Hz}}$) for our protocol is calculated using eq.~(\ref{shot-noise}) and is plotted in solid green, where $L\sim2.0$ cm. The noise floor corresponds to a mirror's differential displacement of about (200 frad/$\sqrt{\mbox{Hz}}$)x(2.0 cm)$\sim$4.0 fm/$\sqrt{\mbox{Hz}}$, and the shot-noise limit to $\sim$1.1 fm/$\sqrt{\mbox{Hz}}$. 

We observed that the system dramatically deviates from the shot-noise limit for low frequencies (<1 Hz). The piezo controller and the function generator feeding the reference signal contributed to slow drift and instability, respectively. In order to evaluate the true performance of the technique to external tilts, all electronics feeding the piezo stack were turned off. The result is shown in dashed red. Such a spectrum corresponds to a 40-hour run taken by combining 240, back-to-back acquired, 10-min runs. For this case, the low-noise preamplifiers were set as 6 dB/oct rolloff low-pass filters at 300 Hz and 20-folded amplification. The acquisition rate on the oscilloscope was 20 Hz~\footnote{Note that the peak value of the reference peak takes the value $\sim$1.6 nrad/$\sqrt{1 \mbox{kHz}}$/5 in the plot since the spectra are shown using the Matlab function \textit{smooth} with a moving average filter of five. The exact value of the peak was used to set the scale of the plot when setting the moving average filter to one (no smoothing). The red curve was taken without a reference signal so it was aligned to the blue curve at 10 Hz when the moving average filter was 1000.}. The noise floor of the device presents a plateau of $\sim$70 prad/$\sqrt{\mbox{Hz}}$ between 2 and 100 mHz for the mirror's frequency. The noise floor goes up to a value of $\sim$7 nrad/$\sqrt{\mbox{Hz}}$ at a mirror frequency of 10 $\mu$Hz. The low-frequency noise is a mixture of different sources, such as thermal fluctuations or gradients, and external vibrations.  

The information of the tilt $\theta$ is carried by the phase difference between the paths, $\phi=\sqrt{2}k_0L\theta$. The right hand scale in Fig.~\ref{fig:spectrum} is the phase spectral density, where the shot-noise limit is given by $\Delta\phi_{SN}\approx13\mbox{ nrad}/\sqrt{\mbox{Hz}}$. The linear response of the tilt sensor was verified down to the applied reference voltage of 321 $\mu$V (lower data point in inset of Fig.~\ref{fig:spectrum}).

Our proposed technique corresponds to the classical optical version of a inverse weak value protocol~\cite{ReviewNori}. 
The quantum description consists of the following steps: 1) A qubit and a continuous degree of freedom (meter) are prepared in a separable state. We use as a qubit the which-path degree of freedom in the interferometer (clockwise $|\lcirclearrowleft\rangle$ and counter-clockwise $|\rcirclearrowleft\rangle$), and as the meter the vertical transverse coordinate of the TEM00 laser mode. The initial global state is given as $|i\rangle\otimes|\Psi\rangle$, where $|i\rangle=(|\lcirclearrowleft\rangle+|\rcirclearrowleft\rangle)/\sqrt{2}$ is the beam's state right after entering the interferometer, and $|\Psi\rangle=(\sqrt{2\pi}\sigma)^{-1/2}\int_{-\infty}^{\infty}dz\,e^{-z^2/4\sigma^2}|z\rangle$. 2) The qubit  and the meter are weakly coupled via the interaction $e^{-ig\hat{\sigma}_3\otimes\hat{z}}$, where $\hat{\sigma}_3$ is the third Pauli's matrix and $g$ is the interaction strength. In our protocol $2g=k$ is the beams' misalignment (effective vertical momentum kick difference between paths) which induces the bi-modal distribution, and $\sigma_3=|\lcirclearrowleft\rangle\langle\lcirclearrowleft|-|\rcirclearrowleft\rangle\langle\rcirclearrowleft|$. 3) Finally, the qubit is post-selected to the state $|f\rangle=(|\lcirclearrowleft\rangle-e^{i\phi}|\rcirclearrowleft\rangle)/\sqrt{2}$, and the shift in the mean of the meter degree of freedom is measured from the events succeeding the post-selection. In our case, the post-selection is done by tracking the dark port of the interferometer where the phase $\phi=\sqrt{2}k_0L\theta$ carries the information of the tilt.


The final state of the meter takes the form,
\begin{eqnarray}
|\Psi_f\rangle&=&\frac{1}{\sqrt{P}}\langle f|e^{-ig\hat{\sigma}_3\otimes\hat{z}}|i\rangle|\Psi\rangle\nonumber\\
&=&\frac{\langle f|i\rangle}{\sqrt{\sqrt{2\pi}\sigma P}}
\int_{-\infty}^{\infty}dz\left[\cos(kz/2)-i\sin(kz/2)\sigma_w\right]e^{-z^2/4\sigma^2}|z\rangle,\nonumber
\end{eqnarray}
where $\sigma_w$ is the weak value of $\sigma_3$ for pre-selection $|i\rangle$ and post-selection $\langle f|$, and $P$ is the probability of post-selection, i.e.
\begin{eqnarray}
\sigma_w&=&\frac{\langle f|\hat{\sigma}_3|i\rangle}{\langle f|i\rangle}=-i\cot(\phi/2),\mbox{ and}\\
P
&\approx&\sin^2(\phi/2)+(k\sigma/2)^2\cos(\phi),
\end{eqnarray}
where we have assumed a weak qubit-meter interaction, $k\sigma\ll1$.

The expectation value of the coordinate $z$ in the post-selection surviving meter events is shifted by the amount,

\begin{equation}
\langle z\rangle=\langle\Psi_f|\hat{z}|\Psi_f\rangle
\approx\frac{2k\sigma^2\sin(\phi)}{4\sin^2(\phi/2)+k^2\sigma^2\cos(\phi)}.
\end{equation}


The metrological technique of our presented experiment is known as inverse weak value amplification~\cite{ReviewNori}. This approach is defined when $k\sigma|\sigma_w|\gg1$ or $\phi/2\ll k\sigma\ll1$. For such a case, the backaction on the meter of the pre- and post-selection process overcomes the one of the weak qubit-meter intermediate interaction. As a result, the input Gaussian distribution for $|\Psi\rangle$ is highly disturbed and a bimodal distribution (as shown in Fig.~\ref{fig:density}) is obtained. The mean of the distribution is given by $\langle z\rangle\approx-(4/k)[\mbox{Im}(\sigma_w)/|\sigma_w|^2]\approx2\phi/k$ (see Eq.~\ref{shift}), and the probability of post-selection takes the form $P\approx(k\sigma/2)^2$. The best known (or most used for parameter estimation) to date of the post-selected weak measurements techniques is weak-value amplification (WVA)~\cite{AAV}. This technique consists in the case where the interaction $k$ is very small and post-selection is used to estimate it. In such a case, where $k\sigma|\sigma_w|\ll1$ or $k\sigma\ll\phi/2\ll1$, the final state for the meter, $|\Psi_f\rangle$, remains as a Gaussian distribution but shifted by an amount $\langle z\rangle\approx-k\sigma^2\mbox{Im}(\sigma_w)\approx2k\sigma^2/\phi$. The probability for success of the post-selection takes the form $P\approx\sin^2{(\phi/2)}$. 

In summary, we have presented an inverse-weak-value amplification metrological technique for precision measurements of ultra-small tilts inside a modified Sagnac interferometer. 
The simplicity and lack of a demanding detection protocol (one split detector is used) of the technique promises advantages over other interferometric techniques. Our best reported result is of a differential displacement noise-floor of  $\sim$4 fm/$\sqrt{\mbox{Hz}}$ for frequencies above 2 Hz. We believe this approach is a good candidate for experiments related to gravity measurements where good performance at low frequencies is required. 

The authors thank Army Research Office (Grant No. W911NF-12-1-0263), Northrop Grumman Corporation, Department of Physics and Astronomy at the University of Rochester, and National Natural Science Foundation of China (NSFC) (11374368)  for financial support.


\end{document}